\documentclass[aps,prl,showpacs,twocolumn,groupedaddress]{revtex4}  
\usepackage{graphicx}  
\usepackage{dcolumn}   
\usepackage{bm}        
\usepackage{amssymb}   

\begin{document}


\hspace{5.2in}
\mbox{FERMILAB-PUB-08-034-E}

\title{\boldmath  
Measurement of the inclusive jet cross section 
in $p \bar{p}$ collisions at $\sqrt{s}=1.96\,{\rm TeV}$ \\}
%
\author{V.M.~Abazov$^{36}$}
\author{B.~Abbott$^{75}$}
\author{M.~Abolins$^{65}$}
\author{B.S.~Acharya$^{29}$}
\author{M.~Adams$^{51}$}
\author{T.~Adams$^{49}$}
\author{E.~Aguilo$^{6}$}
\author{S.H.~Ahn$^{31}$}
\author{M.~Ahsan$^{59}$}
\author{G.D.~Alexeev$^{36}$}
% ...

\affiliation{\vspace{0.1 in}(The D\O\ Collaboration)\vspace{0.1 in}}
% ...
\affiliation{$^{6}$University of Alberta, Edmonton, Alberta, Canada,
                Simon Fraser University, Burnaby, British Columbia, Canada,
                York University, Toronto, Ontario, Canada, and
                McGill University, Montreal, Quebec, Canada}
% ...
\affiliation{$^{29}$Tata Institute of Fundamental Research, Mumbai, India}
% ...
\affiliation{$^{31}$Korea Detector Laboratory, Korea University, Seoul, Korea}
% ...
\affiliation{$^{36}$Joint Institute for Nuclear Research, Dubna, Russia}
% ...
\affiliation{$^{49}$Florida State University, Tallahassee, Florida 32306, USA}
% ...
\affiliation{$^{51}$University of Illinois at Chicago,
                Chicago, Illinois 60607, USA}
% ...
\affiliation{$^{59}$Kansas State University, Manhattan, Kansas 66506, USA}
% ...
\affiliation{$^{65}$Michigan State University,
                East Lansing, Michigan 48824, USA}
% ...
\affiliation{$^{75}$University of Oklahoma, Norman, Oklahoma 73019, USA}
% ...
\date{\today}
\begin{abstract}
We report on a measurement of the inclusive jet cross section in $p \bar{p}$
collisions at a center-of-mass energy $\sqrt s=$1.96~TeV using data collected by the
D0 experiment at the Fermilab Tevatron Collider corresponding to an
integrated luminosity of 0.70~fb$^{-1}$. 
 The data cover jet transverse momenta
from
50~GeV to 600~GeV and jet rapidities in the range -2.4 to 2.4. Detailed studies of  
correlations
between systematic uncertainties in transverse momentum and rapidity are
presented, and
the cross section measurements are found to be in good agreement with next-to-leading order QCD
calculations. 
\end{abstract}

\maketitle
The measurement of the cross section for the inclusive production of jets in 
hadron collisions 
provides stringent tests of quantum chromodynamics (QCD).
When the transverse momentum ($p_T$) of the jet with respect to the
beam axis is large,
contributions from long-distance processes 
are small and the production of jets can be calculated
in perturbative QCD (pQCD).
The inclusive jet cross section in $p\bar{p}$ collisions
at large $p_T$ provides one of the most direct probes of physics at 
small distances. In particular, it is
directly sensitive to the strong coupling constant ($\alpha_s$) and the parton distribution
functions (PDFs)
of the proton. Additionally, it can be used to set constraints on the
internal structure of quarks~\cite{contact}.
Deviations from  pQCD predictions at large $p_T$
can indicate new physical phenomena
not described by the standard model of particle physics. A measurement over the widest possible 
rapidity range provides 
simultaneous sensitivity to the PDFs as well as new phenomena 
expected to populate mainly low rapidities. These data will have a strong 
impact on physics at the CERN Large Hadron Collider (LHC)
where searches for new particles and higher dimensions suffer
from poor knowledge of PDFs~\cite{newphenomena}.

In this Letter, we report on a 
measurement from the D0 experiment of the inclusive jet cross section
in $p\bar{p}$ collisions at a center-of-mass of $\sqrt{s}=1.96$~TeV. The data
sample, collected with the D0 detector during 2004-2005 in Run~II of
the Fermilab Tevatron Collider, corresponds to an integrated
luminosity of ${\cal L}=0.70$~fb$^{-1}$~\cite{lumi}.  The increased $p \bar{p}$ 
center-of-mass
energy between Run~I ($\sqrt{s}=1.8$~TeV) and Run~II
leads to significant increase in the cross section at large $p_T$ --- a
factor of three at $p_T \sim 550$~GeV. The cross section is presented in six
bins of jet rapidity ($y$), extending out to $|y|=2.4$, as a function of
jet $p_T$ starting at $p_T=50$~GeV, and provides the largest data
set of the inclusive jet spectra at the Tevatron with the smallest
experimental uncertainties to date. The measurement also extends
earlier inclusive jet cross section measurements by the CDF and D0
collaborations~\cite{cdf,d0_runI} and improves the systematic uncertainties
compared to previous measurements by up to a factor of two over a range of rapidity
up to 2.4 at high $p_T$.

The primary tool for jet detection is the
finely segmented liquid-argon and uranium calorimeter that
has almost complete solid angular coverage~\cite{d0det}. The central calorimeter (CC) 
covers the pseudorapidity 
region $|\eta|<1.1$
and the two endcap calorimeters (EC) extend the coverage up to $|\eta| \sim 4.2$.
The intercryostat region (ICR) between the CC and EC contains
scintillator-based detectors that supplement the coverage of the 
calorimeter.
The Run~II iterative seed-based cone jet algorithm including
mid-points~\cite{run2cone} with cone radius ${\cal R}=\sqrt{(\Delta y)^2 +
(\Delta \phi)^2}=$ 0.7 
in rapidity $y$ and azimuthal angle $\phi$ is used to
cluster energies deposited in calorimeter towers. The same algorithm is used for partons in the
pQCD calculations.
The binning in jet $p_T$ is commensurate with the
measured $p_T$ resolution.

Events are required to satisfy jet trigger requirements.  
Only jets above a given $p_T$
threshold are kept by the highest level trigger (L3). The cross section
is corrected for jet trigger inefficiencies (always below 2\%) determined using 
an independent sample of muon triggered
events.

The jet $p_T$ is corrected for the energy
response of the calorimeter, energy showering in and out the jet cone, 
and additional energy from event pile-up
and multiple proton interactions.
After applying these corrections, the jet four momentum is 
given at the particle level, which means that they represent the real energy
of the jet made out of the stable
particles resulting from
the hadronization process following the hard $p \bar{p}$ interaction..
The electromagnetic part of the calorimeter
is calibrated using $Z \rightarrow e^+e^-$ events~\cite{zee}. The jet 
response for the region $|\eta|<0.4$
is determined using the momentum imbalance in
$\gamma +$jet events. 
The $p_T$ imbalance in dijet events with one jet in 
$|\eta|<0.4$ and the other anywhere in $\eta$ is used to intercalibrate 
the jet response in $\eta$, as a function of jet $p_T$. 
Jet energy scale corrections are typically  $\sim$$50\%$ ($20\%$)
of the jet energy at $50$ ($400$)~GeV.
Further corrections due to the difference in response between
quark- and gluon-initiated jets are computed using the {\sc pythia}~\cite{PYTHIA}
event generator, passed through a  
{\sc geant}-based~\cite{geant} simulation of the detector response. 
These corrections amount to $\sim+4\%$
at jet energies of $50$~GeV and $\sim-2\%$ at 400~GeV in the CC.
The relative uncertainty of the jet $p_T$ calibration ranges from $1.2\%$ at $p_T\sim
150$~GeV to $1.5\%$ at $500$~GeV in the CC, and 1.5--2\% in the ICR and EC .

The position of the $p\bar{p}$ interaction is reconstructed using a
tracking system consisting of silicon
microstrip detectors and scintillating fibers located inside a solenoidal magnetic
field of $2\,\text{T}$ \cite{d0det}. The position of the vertex along
the beamline is required to be within
$50$\,cm of the detector center. 
The signal efficiency of this requirement is $93.0 \pm 0.5\%$.
A requirement is placed on the missing transverse energy
in the event, computed as the transverse
component of the vector sum of the
momenta in calorimeter cells, to
suppress the cosmic ray background and is
$>99.5\%$ efficient for signal. 
Requirements on characteristics of shower development 
are
used to remove the remaining background due to electrons, photons, and detector
noise that mimic jets. The efficiency for these requirements is $>99\%$
($>97.5\%$ in the ICR). After all these requirements, the background is
$<0.1$\% in our sample.

The D0 detector
simulation provides a good description of jet properties
including characteristics of the shower development. 
The correction to the jet cross section for muons and
neutrinos, not reconstructed within jets, is determined using {\sc
pythia} and is $2\%$, independent of $p_T$ and $y$.
The corrections for jet migration between bins in $p_T$ and $y$ due
to finite resolution in energy and position are determined in an unfolding
procedure, based on the experimental $p_T$ and $y$ resolutions.
The jet $p_T$ resolution is obtained using the $p_T$
imbalance in dijet events and is found to decrease from
13\% at $p_T \sim 50$~GeV to 7\% at $p_T \sim 400$~GeV in both the 
CC and the EC. The 
resolution in the ICR is 16\% at $p_T \sim 50$~GeV decreasing to
11\% at $p_T \sim400$~GeV.
The method to unfold the data uses
a four-parameter ansatz function~\cite{Ansatz} 
to parametrize 
the $p_T$ dependence of the jet cross section convoluted with the measured $p_T$ resolution
and fitted to the experimental data.

The unfolding corrections vary between $20\%$ at a jet $p_T\sim 50$~GeV and
$40\%$
at $400$~GeV in the CC. In the EC and the ICR, the corrections are less than
$20\%$ at 
$p_T\sim 50$~GeV, but increase to $80\%$ at the largest $p_T$ and $y$. Bin 
sizes in $p_T$ and $y$ 
are chosen to minimize migration corrections
due to the experimental resolution.
The $y$ resolution is better 
than $0.05$ ($0.01$) for jets with
$p_T\sim 50$~GeV ($400$~GeV), and leads to a migration
correction less than 2\% in most bins, and 10\% in the highest $y$
bin.

\begin{figure}
\includegraphics[scale=0.42]{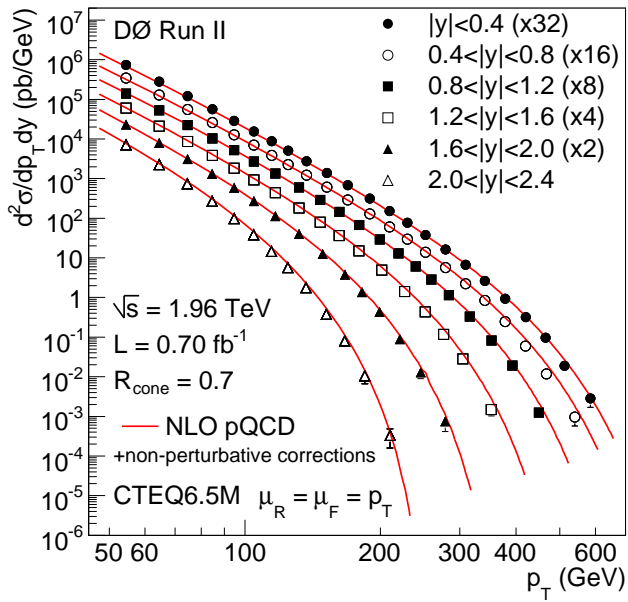}
\caption{\label{fig1} 
The inclusive jet cross section as a function of jet $p_T$ in six $|y|$ bins.
The data points are multiplied by 2, 4, 8, 16, and 32
for the bins $1.6<|y|<2.0$,
$1.2<|y|<1.6$, $0.8<|y|<1.2$, $0.4<|y|<0.8$, and $|y|<0.4$, respectively. 
}
\end{figure}

\begin{figure*}
\includegraphics[scale=0.78]{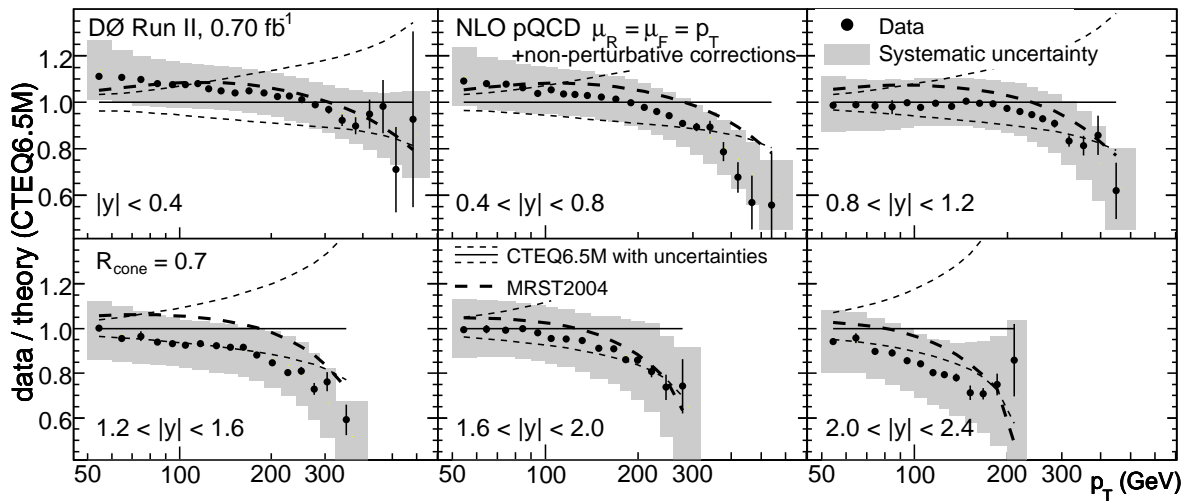}
\caption{\label{fig2} 
Measured data divided by theory for the inclusive jet cross section 
as a function of jet $p_T$ in six $|y|$ bins.
The data systematic uncertainties
are displayed by the full shaded band.
NLO pQCD calculations,
with renormalization and factorization scales set to jet $p_T$ using
the CTEQ6.5M PDFs and including non-perturbative corrections, are
compared to the data. The CTEQ6.5 PDF uncertainties are shown as small
dashed lines and the predictions with MRST2004 PDFs as large dashed
lines. The theoretical scale uncertainty,
obtained by varying the factorization and
renormalization scales between $\mu_R = \mu_F = p_T /2$ and $\mu_R = \mu_F = 2
p_T $,
is typically 10--15\%
}
\end{figure*}

The results of the inclusive jet cross section measurement corrected
to the particle level
are displayed in Fig.~1 in 
six $|y|$ bins as a function of $p_T$. The cross section extends over 
more than eight orders of
magnitude from $p_T=50$~GeV to $p_T>600$~GeV.  
Perturbative QCD predictions to next-to-leading order (NLO) in $\alpha_S$, 
computed using the 
{\sc fastNLO} 
program~\cite{fastnlo} (based on
{\sc nlojet++}~\cite{nlojet})
and the PDFs from CTEQ6.5M~\cite{cteq6}, are
compared to the data.
The renormalization and factorization scales ($\mu_R$ and $\mu_F$) are set
to the individual jet $p_T$.
The theoretical uncertainty, determined by
changing $\mu_R$ and $\mu_F$ between $p_T/2$
and $2 p_T$, is of the order of 10\% in all bins.
The predictions are corrected for non-perturbative contributions due to
the underlying event and 
hadronization computed by
{\sc pythia} 
with the CTEQ6.5M PDFs, the QW tune~\cite{qwtune}, and the two-loop formula for $\alpha_S$.
These non-perturbative 
corrections to theory
extend from $+10\%$ to $+20\%$ at $p_T \sim 50$~GeV
between $|y|<0.4$ and $2.0 < |y| < 2.4$. The corrections are of order 
$+5\%$ for $p_T \sim 100$~GeV, and smaller than $+2\%$
above 200~GeV.

The ratio of the data to the theory is shown in Fig.~2. 
The dashed lines
show the uncertainties due to the different PDFs 
coming from the CTEQ6.5 parametrizations.
The predictions from MRST2004~\cite{mrst2004} are displayed by the large 
dashed line. 
In all $y$ regions, the
predictions agree well with the data. There is a tendency for the data to be lower than the
central CTEQ prediction --- particularly at very large $p_T$ --- but
they lie mostly within the CTEQ PDF uncertainty band. The $p_T$ dependence of the data
is well reproduced by the MRST parametrization whose systematic
uncertainty is slightly smaller
than that from the CTEQ parametrization. The experimental
systematic uncertainty is
comparable to the PDF uncertainties. The theoretical scale uncertainty,
obtained by varying the factorization and
renormalization scales between $\mu_R = \mu_F = p_T /2$ and $\mu_R = \mu_F = 2
p_T $,
is typically 10--15\%.
In most bins, the experimental uncertainties are of the same order as the
theoretical uncertainties.
Tables of the cross sections together with
their uncertainties are given in Ref.~\cite{table}.

Correlations between systematic uncertainties are studied in detail 
to increase the value of these data in future PDF fits~\cite{table} and their impact on
LHC physics in particular. 
Point-to-point correlations in $p_T$ and
$y$ are provided for the 24 sources of systematic uncertainty. 
The relative uncertainties in percent on the
cross section measurement are shown in Fig.~3 for the five most significant
sources of systematic uncertainty in
$|y|<0.4$ and $2.0<|y|<2.4$. 
The luminosity uncertainty of 6.1\%, fully correlated in
$p_T$ and $y$, is not displayed in Fig.~3.
The other $y$ bins have
similar correlations in shape and values between these two extreme bins.
The total uncorrelated uncertainty is
$<3\%$ in the CC, 
and $<15\%$ in the EC. 

The two largest systematic uncertainties are due to
the electromagnetic energy scale obtained from $Z\rightarrow e^+e^-$ events~\cite{zee}, 
and the photon energy scale in the CC obtained using
the difference in the
calorimeter response between photons and electrons in the detector
simulation. The uncertainty
on the photon energy scale is mainly due to the limited
knowledge of the amount of dead material in
front of the calorimeter and from the physics modeling of electromagnetic
showers in the {\sc geant}-based~\cite{geant} simulation. These two
contributions to the jet cross section uncertainty are $\sim 5\%$ in the CC and $5-15\%$ 
in the EC. 

The large-$p_T$ extrapolation of jet energy scale is determined using the 
detector simulation
with the single-pion response tuned to $\gamma$+jet data. 
The uncertainty rises to $12\%$ ($30\%$) in 
the CC (EC), and is dominated by the uncertainty in the jet
fragmentation, estimated by comparing the fragmentation models in {\sc pythia}
and {\sc herwig}~\cite{HERWIG}. The uncertainty in $\eta$ intercalibration
corresponds to systematic uncertainties associated with the procedure to equalize 
the calorimeter response in
different regions of $\eta$ in dijet events. 
These systematic uncertainties are negligible in the CC because the
$\eta$ dependent response is calibrated with respect to the CC, 
but extend up to 25\% in the EC. Finally, systematic uncertainties associated
with showering effects, due primarily to the modeling of
the hadronic shower development in the detector and differences between {\sc pythia}
and {\sc herwig}, range from $3\%$ at low $p_T$ to $7\%$ ($15\%$) at
large $p_T$ in the CC (EC).

To show the potential impact of using 
point-to-point uncertainty correlations in jet $p_T$ and $y$ on PDF
determination, we give in Fig.~3 the uncorrelated and
total systematic uncertainties as a function of jet $p_T$ as a percentage of the jet cross section
measurement. The total uncorrelated uncertainties are less than $15\%$ and $25\%$ of the full
uncertainties in the CC and EC respectively. The
full systematic uncertainties are similar in size to the PDF uncertainties 
(Fig.~2) and the detailed analysis of the correlations which have been performed
will make it possible to further constrain the 
PDFs. Knowledge of these correlations 
is especially important for constraining the PDFs in NNLO pQCD
fits where the uncertainties due to the dependence on the choice
of the renormalization and factorization scales are smaller. The point-to-point
correlations for the 24 different sources of systematic
uncertainties are given in Ref.~\cite{table}.

In conclusion, the measured inclusive jet cross section corrected
for experimental effects to the particle level
in $p\bar{p}$ collisions at $\sqrt{s}=$ 1.96 TeV with ${\cal
L}=0.70$~fb$^{-1}$ is
presented
for six $|y|$ bins as a function of jet $p_T$, substantially extending the kinematic reach 
and improving the precision of existing 
inclusive jet measurements.
NLO pQCD calculations with CTEQ6.5M 
or MRST2004 PDFs agree with
the data and favor the lower edge of the CTEQ6.5 PDF uncertainty band at large
$p_T$ and the shape of the $p_T$ dependence for MRST2004.  
A full analysis of correlations between sources of systematic uncertainty is
performed, increasing the potential impact
of these data in global PDF fits and on new phenomena searches at the 
LHC.

\begin{figure}
\includegraphics[scale=0.42]{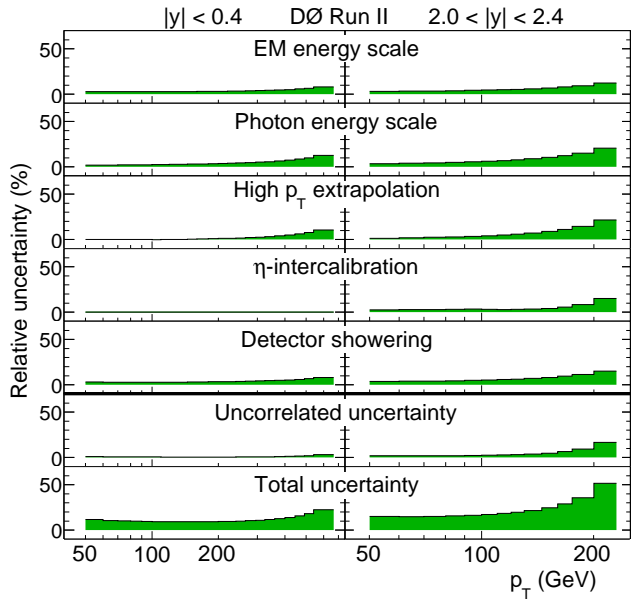}
\caption{\label{fig3} 
Correlated uncertainties for $|y|<0.4$ and
$2.0<|y|<2.4$ as a function of jet $p_T$. The five largest systematic
uncertainties are shown together with uncorrelated and total uncertainties,
computed as a sum in quadrature of all sources.
}
\end{figure}

%
We thank the staffs at Fermilab and collaborating institutions, 
and acknowledge support from the 
DOE and NSF (USA);
CEA and CNRS/IN2P3 (France);
FASI, Rosatom and RFBR (Russia);
CAPES, CNPq, FAPERJ, FAPESP and FUNDUNESP (Brazil);
DAE and DST (India);
Colciencias (Colombia);
CONACyT (Mexico);
KRF and KOSEF (Korea);
CONICET and UBACyT (Argentina);
FOM (The Netherlands);
STFC (United Kingdom);
MSMT and GACR (Czech Republic);
CRC Program, CFI, NSERC and WestGrid Project (Canada);
BMBF and DFG (Germany);
SFI (Ireland);
The Swedish Research Council (Sweden);
CAS and CNSF (China);
and the
Alexander von Humboldt Foundation.

\end{document}